\documentclass[graybox]{svmult}
\usepackage{amsmath}
\usepackage{amssymb}
\usepackage{mathptmx}       % selects Times Roman as basic font
\usepackage{helvet}         % selects Helvetica as sans-serif font
\usepackage{courier}        % selects Courier as typewriter font
\usepackage{type1cm}        % activate if the above 3 fonts are
\usepackage{makeidx}         % allows index generation
\usepackage{graphicx}        % standard LaTeX graphics tool
\usepackage{multicol}        % used for the two-column index
\usepackage[bottom]{footmisc}% places footnotes at page bottom

\makeindex             % used for the subject index
\begin{document}

\title*{Parameter selection in dynamic contrast-enhanced magnetic resonance tomography}
\author{Niinim\"aki, K and Hanhela, M and Kolehmainen, V}
\maketitle

\abstract{In this work we consider the image reconstruction problem of
  sparsely sampled dynamic contrast-enhanced (DCE) magnetic resonance
  imaging (MRI). DCE-MRI is a technique for acquiring a series of MR
images before, during and after intravenous contrast agent
administration, and it is used to study microvascular structure and
perfusion. To overcome the ill-posedness of the related spatio-temporal
inverse problem, we use regularization. In regularization one of the main problems is how to determine the regularization parameter which controls the balance between data fitting term and regularization term. Most methods for selecting this parameter require the computation of a large number of estimates even in stationary problems. In dynamic imaging, the parameter selection is even more time consuming since separate regularization parameters are needed for the spatial and temporal regularization functionals. In this work, we study the possibility of using the S-curve with DCE-MR data.
We select the spatial regularization parameter using the S-curve, leaving the temporal regularization parameter as the only free parameter in the reconstruction problem. In this work, the temporal regularization parameter is selected manually  
by computing reconstructions with several values of the temporal regularization parameter.
}

\section{Introduction}
\label{sec:intro}
Dynamic contrast-enhanced MRI (DCE-MRI) is an imaging method which is used to study microvascular structure and tissue perfusion. The method has many applications including blood-brain-barrier assessment after acute ischemic stroke \cite{Merali2017,Villringer2017} and treatment monitoring of breast cancer \cite{Martincich2004,Pickles2005} and glioma \cite{Piludu2015}. The operation principle of DCE-MRI is to inject a bolus of gadolinium based contrast agent into the blood stream, and acquire a time series of MRI data with a suitable $T_1$-weighting to obtain a time series of 2D (or 3D) images which exhibit contrast changes induced by concentration changes of the contrast agent in the tissues.

The analysis of the contrast agent dynamics during imaging requires
high resolution in space and in time; the high temporal resolution is necessary to measure when the contrast is passing
through the artery and it is used for determining the subject specific Arterial Input function (AIF), and the high
spatial resolution is necessary to adequately capture boundaries of perfused tissues.
In many cases, sufficient time resolution can only be obtained by utilizing an imaging protocol during
which only partial k-space sampling can be obtained for each image in the time series.
However,  acquiring less samples than required by
the Nyquist criterion makes the image reconstruction problem {\em ill-posed} (and non-unique), 
causing artifacts and deterioration of the image quality when conventional reconstruction methods such as 
inverse Fourier transform or re-gridding are employed.

According to the theory of compressed sensing (CS)
\cite{Cand`es2006a,Cand`es2006,Cand`es2006b}, images that have a
sparse representation can be recovered from undersampled
measurements of a linear transform, i.e. sampling rate below Nyquist
rate, using appropriate nonlinear reconstruction algorithms and appropriate 
(random) sampling of the data space. The compressibility 
% (sparse in some transformation domain) 
of MR images  and the fact that MR scanner measures 
samples of a linear transform of the unknown image  (the k-space samples can be mathematically considered as  Fourier coefficients) suggest that the idea of CS is applicable to MR imaging, offering thus a
potentially significant scan time reduction without sacrificing the image quality. 
Since the seminal work of Cand\`{e}s, Romberg and Tao in 2006, CS has
been extensively applied to MRI.  In 2007 CS was applied to MRI in \cite{lustig2007_sparseMRI} and in 2018 it received FDA
approval for clinical use.

The undersampling of the k-space for speeding up the dynamic MRI data acquisition 
can in principle be done in many different ways. However, for many applications, the low frequency features that are present in the center of the k-space are of importance. In
\cite{Zong2014}, for example, it was demonstrated that center weighted
random sampling patterns were preferable to purely random sampling of
the k-space within the CS approach. 
%In this work, we consider radial sampling of the k-space. 
Radial sampling has the advantage that the center of the k-space is sampled
densely, even when the sampling (i.e., number of radial spokes) is remarkably reduced. 

In this paper, we consider image reconstruction problem of DCE-MRI with sparsely sampled golden angle radial data, where the angle of subsequent spokes is $\sim 111.25^\circ$. The number of measurement spokes used for reconstructing a single time frame is chosen to be a Fibonacci number, which was shown to be an optimal choice in \cite{Winkelmann2007}. This type of sparse sampling between the time frames results in each time frame having different spokes, and eventually the spokes cover the whole k-space.

We also combine the Golden Angle (GA) sampling with concentric squares sampling. This sampling strategy resembles the linogram method \cite{Edholm1987} developed for computed tomography imaging, but the angles of subsequent spokes were chosen according to the golden angle method as opposed to the angles being equidistant in $\tan\theta$ in linogram sampling. Unlike the conventional radial sampling pattern with spokes of equal length, the concentric squares sampling strategy also covers the corners of the k-space. The sampling pattern therefore also collects information of the high frequencies in the corners of the k-space, which leads to a reduction of artifacts originating from the lack of sampling in the corners.

To overcome the ill-posedness of our image reconstruction problem, we use a variational framework. Thus we solve the following optimization problem

\begin{equation}
  \label{eq:regularizatio_general}
  \hat{f} = \arg \min D(f, m) + \alpha S(f) + \beta T(f),
\end{equation}
where $f=\{f_1,f_2, \ldots ,f_{N_t} \} $ denotes the image sequence, $m =\{m_1,m_2, \ldots ,m_{N_t} \}$ the data sequence, $N_t$ the number of time frames, $D$ the data-fitting term, $S$ the spatial
regularization, $T$ the temporal regularization and $\alpha$ and $\beta$
are the spatial and temporal regularization parameters, respectively. The selection of the
regularization parameters is crucial in terms of resulting image
quality. There exists several proposed parameter selection methods,
but most of them require the computation of a large number of reconstructions with varying parameters. Having to fix two regularization parameters makes the selection even more time consuming.  

In this work, we propose to
%for automatic selection of the regularization parameter, we propose to
 use the S-curve method \cite{hamalainen2013,Niinimaki2013,Kolehmainen2012,niinimaki2015}
 for automatic selection of the spatial regularization parameter $\alpha$. The idea in the S-curve method is to
select the regularization parameter so that the reconstruction has {\em a priori} defined level of sparsity in the chosen transformation domain.  In DCE-MRI, a reliable {\em a priori} estimate for the sparsity level can be extracted from an anatomical MRI image which is based on full-sampling of the k-space and is always taken as part of the MRI measurement protocol but is usually used only for visualization purposes. 
For the selection of the spatial regularization parameter $\alpha$, we employ one time frame of the GA data from the baseline measurement before the contrast agent administration. %Here we used the first frame of the baseline. 
After fixing the spatial
regularization parameter, we compute dynamic reconstructions with several values of parameter $\beta$ and select a suitable temporal regularization parameter manually. Furthermore we study the performance of three different temporal regularization functionals, namely temporal smoothness, temporal total variation and total generalized variation. %such that the resulting %reconstruction  gives the best image fidelity measures.

The proposed method is evaluated using simulated GA DCE-MRI data from 
a rat brain phantom. 
The results are compared to re-gridding approach, which is the most widely used non-iterative algorithm for reconstructing images from non-Cartesian MRI data. Our re-gridding method was developed in IR4M UMR8081, CNRS, Universit\'e Paris-Sud using Matlab$^{\textregistered}$. This re-gridding approach does not need additional density correction and it was first used in \cite{Kusmia2013}, see also \cite{iddn_codeGG}.
%For more information on the code, we refer to 

\section{Image reconstruction in radial DCE-MRI}

\subsection{Forward problem}

The forward problem in 2D MRI  
%represented using the following integral equation
can be 
modelled for most measurement protocols 
by the Fourier transform
\begin{equation}
  \label{eq:directMRIprob}
m(k_x, k_y) = \int_{\Omega} f(x,y)  e^{-i2\pi (k_x x + k_y y)} \mathrm{d}x  \mathrm{d}y,
\end{equation}
where $\Omega$ is the image domain $f(x,y)$  is the unknown image,  $m(k_x, k_y)$ is the measured data, and $k_x, k_y$ denotes the k-space trajectories. 
In the discrete framework, the Fourier transform is typically approximated with the multidimensional FFT when using cartesian k-space trajectories and with the non-uniform FFT when using non-cartesian k-space trajectories.  

In this work, we consider non-uniform k-space trajectories and approximate the Fourier transform by the non-uniform fast Fourier transform (nuFFT) operator \cite{FesslerSutton}. We discretize our functions as follows; temporal direction is divided into a sequence of $N_t$ (vectorized) images  
$f = \{ f_1, f_2, \ldots, f_{N_t} \}$ and data vectors
$m = \{ m_1, m_2, \ldots, m_{N_t}\}$, where each
%and data frames, each denoted by vectors 
$f_t \in \mathbb{C}^{N_p}$ 
and
$m_t \in \mathbb{C}^M$, respectively. The number of data per frame $M$ is equal to the number of GA spokes per frame times the number of samples per spoke. The number of image pixels is
$N_p = N \times N$.
Thus, using nuFFT we re-write (\ref{eq:directMRIprob}) in a discretized form at time $t$ as
\begin{equation}
  \label{eq:directNUFFTmodel}
  m_t = A_t f_t + \epsilon_t, \quad t = 1, \ldots , N_t,
\end{equation}
$A_t =P_t \mathcal{F}S_t$, where $P_t$ is an interpolation matrix between Cartesian k-space and non-cartesian k-space, $\mathcal{F}$ is the 2D FFT operation and $S_t$ is a scaling matrix. 

\subsection{Inverse problem of dynamic image reconstruction}

The dynamic inverse problem related to the equation (\ref{eq:directNUFFTmodel}) is: given measurement time series $m =\{m_1,m_2, \ldots ,m_{N_t} \}$ and the associated k-space trajectories, solve the unknown images $f =\{f_1,f_2, \ldots ,f_{N_t} \}$. 
n. 
To recover $f$ from $m$, we define the inverse problem 
as the optimization problem 
\begin{equation}
  \label{eq:regularizedGeneral}
  \hat{f} = \arg \min \left \{ \sum_{t=1}^{N_t}  \Vert A_t f_t - m_t \Vert_2^2 + \alpha S(f) + \beta T(f) \right \},
\end{equation}
where $S(f)$ denotes the spatial regularization functional, $\alpha$ the spatial regularization parameter, $\beta$ the temporal regularization parameter and $T(f)$ the temporal regularization functional.

In this work, we study the applicability of S-curve method for selecting the spatial regularization parameter
$\alpha$. Once $\alpha$ has been fixed, the temporal regularization parameter is then selected manually by computing estimates with different values of $\beta$. 
Our minimization problem is based on $L_2$-data fidelity term
for the measurement model and we use spatial total variation regularization for promoting sparsity of the gradients of each image
\cite{Rudin1992}. Furthermore we study the performance of three different temporal regularization functionals for promoting temporal regularity of the image
series. Our spatio-temporal image reconstruction problem thus writes

\begin{equation}
  \label{eq:spatio_temporal_regularizationL2TV_tReg}
  \hat{f} = \arg \min \left \{ \sum_{t=1}^{N_t} \left( \Vert A_t f_t - m_t
    \Vert_2^2 + \alpha \Vert \nabla f_t \Vert_{2,1} \right) + \beta T(f) \right \},
\end{equation}
where the isotropic 2D spatial total variation norm for complex valued
image $f_t$ is defined by
\begin{equation}
\label{eq:isotropicTVspat}
  \Vert \cdot \Vert_{2,1} = \sum_{k=1}^{N} \sqrt{(Re(D_x f_{t,k}))^2+ (Re(D_y f_{t,k}))^2+(Im(D_x f_{t,k}))^2+(Im(D_y f_{t,k}))^2},
\end{equation}
where $Re$ and $Im$ denoting the real and imaginary parts of $f_t$
respectively and $D_x$ and $D_y$ the discrete forward first
differences in horizontal and vertical directions, respectively.

\subsubsection{Temporal regularization 1: Temporal smoothness (TS)} 
%$L_2$ norm of forward first differences in time}

The temporal smoothness regularization (hereafter referred as TS) is defined as the $L_2$ norm of forward first differences in time:
%also called
%temporal smoothness model (and from hereafter referred as TS), is as follows
\begin{equation}
  \label{eq:tReg_fdiffL2}
  T(f) = \sum_{t=1}^{N_t-1} \Vert f_{t+1} - f_t \Vert_2^2.
\end{equation}
This model promotes smooth slowly changing signals, and it has been
used in \cite{Adluru2007} for radial DCE myocardial perfusion imaging. TS regularization was compared with temporal TV regularization in
the same application in \cite{Adluru2008}.

\subsubsection{Temporal regularization 2 : Temporal total variation (TV)} 

Temporal total variation (hereafter referred as TV) is defined by the $L_1$ norm of the forward first differences in time: %accordding to
\begin{equation}
  \label{eq:tReg_fdiffL1}
  T(f) = \sum_{t=1}^{N_t-1} \Vert f_{t+1} - f_t \Vert_1.
\end{equation}

The temporal total variation model promotes sparsity of the time derivative
of the pixel signals, being a highly feasible regularization model for reconstruction of piece-wise regular signals which may exhibit large jumps. The smoothed form of temporal total variation was used in \cite{Adluru2009} for multislice myocardial perfusion imaging.

\subsubsection{Temporal regularization 3: Total generalized variation (TGV)} 

The total generalized variation model \cite{Bredies2010} is a
total variation model that is generalized to higher order differences. Here
we use the second-order total generalized variation, which in the discrete
1-dimensional form is of the form
\begin{equation}
  \label{eq:tReg_tgv}
  T(f) = \sum_{t=1}^{N_t-1} \min_v \Vert f_{t+1}-f_t  - v_t \Vert_1 +
  \gamma \Vert v_{t+1} - v_t \Vert_1,
\end{equation}
where $v$ is an auxiliary vector and $\gamma$ is a parameter, which balances the first and second order terms, and is set to $\sqrt{2}$ here.

This functional balances between minimizing the first-order and second-order
differences of the signal. The difference with TV regularization is most
clear in smooth regions where piecewise linear solutions are favored over the
piecewise constant solutions of TV.  From hereafter this temporal regularization is referred as TGV.

TGV was first used in MRI as a spatial prior in \cite{Knoll2011}, and it has also been used in DCE-MRI as a temporal prior in \cite{Wang2017}, where different temporal priors were compared in cartesian MRI of the breast.

\subsection{Regularization parameter selection}

\subsubsection{Spatial regularization parameter selection}
\label{sec:Scurve}

The spatial regularization parameter is selected using the S-curve
method, originally proposed in
\cite{hamalainen2013,Niinimaki2013,Kolehmainen2012,niinimaki2015}, but here modified for TV
regularization. %The S-curve method with TV regularization  is the following.

Assume that we have an {\it a  priori} estimate  $\hat{S}$ for the
total variation norm of the unknown function. In practice we can use an anatomical image of the same slice in order to obtain a reliable estimate for $\hat{S}$. Such an anatomical image is  practically always acquired as part of the DCE MRI acquisition experiment but usually only used for visualization purposes. However, if such an anatomical image was not acquired, we could, in case of GA acquisition, estimate the expected sparsity level from a conventional reconstruction of a long sequence of baseline data taken before the contrast agent injection. Or the anatomical image could be estimated from the entire data-acquisition similarly as the composite image in \cite{Mistretta2006}.

Now, given the estimate $\hat{S}$ we select the regularization
parameter $\alpha$ using the S-curve method as follows

\begin{itemize}
\item [1)] Take a sequence of regularization parameters $\alpha$ ranging on the interval $[0, \infty]$
 such that
$$
 0 < \alpha^{(1)} < \alpha^{(2)} < \cdots < \alpha^{(L)} < \infty .
$$
\item [2)] Compute the corresponding estimates $\hat{f_1} (\alpha^{(1)}),
 \ldots, \hat{f_1} (\alpha^{(L)})$. \\
 With DCE-MRI data, reconstructions $ \hat{f_1} (\alpha^{(\ell)})$ are computed as follows; we take the data that correspond to the first time frame, i.e. $m_1$ which has number of elements equal to the number of GA spokes per frame times the number of points per spoke. 
 %Now our optimization problem becomes stationary and 
 We reconstruct $f_1$ for given value $\alpha^{(\ell)}$ %the corresponding estimates $\hat{f_1}$ 
 by 
 %minimizing the following functional
 \begin{displaymath}
     \hat{f_1}(\alpha^{(\ell)}) = \arg \min_{f_1} \{ \|A_1 f_1 -  m_1 \|_2^2 + \alpha^{(\ell)} \| \nabla f_1 \|_{2,1} \}.
 \end{displaymath}
Here it is important that $\alpha^{(1)}$ is taken to be so
 small that the problem is under regularized and the corresponding reconstruction $\hat{f_1}(\alpha^{(1)})$ results to a very noisy image with a big TV-norm value and
 $\alpha^{(L)}$ is taken so large that the problem is over regularized and TV norm of reconstruction
$\hat{f_1}(\alpha^{(L)})$ is very close to zero. 
\item[3)] Compute the TV-norms of the recovered estimates $\hat{f_1} (\alpha^{(\ell)}), \quad \ell = 1,\ldots,L$.
\item [4)] Fit a smooth
 interpolation curve to the data $\{ \alpha^{(\ell)},S(\alpha^{(\ell)}), \quad \ell = 1,\ldots,L\}$ and use the interpolated sparsity curve  to find the value of $\alpha$ for which $S(\alpha) = \hat{S}$. For the interpolation we use Matlab's$^{\textregistered}$ {\it interp} function and we interpolate our original S-curve to a more dense discretization of the regularization parameter $\alpha$. 
 %a dozen values to several hundreds of values.
\end{itemize}

\subsection{Temporal regularization parameter selection}

Once the spatial regularization parameter $\alpha$ has been fixed using the S-curve, the temporal regularization parameter $\beta$ can be tuned by computing estimates with different values of $\beta$ and selecting a suitable value manually, for example, by visual assessment of the results.

In this work, we compute the results with three different temporal regularization models using simulated measurement data. 
%we select the temporal regularization parameter $\beta$ manually. However, 
Since we consider a simulated test case where a ground truth is available, we select an optimal value of $\beta$ for each temporal regularization model by selecting the value of $\beta$
which produces the reconstruction with the smallest root mean square error (RMSE).
The RMSE values were calculated separately  for three regions; tumor, vascular region and the rest of the image domain. The RMSEs of different ROIs were then used to define a joint RMSE as
\begin{equation}
  \label{eq:RMSEjoint}
  RMSE_{joint} = \sqrt{RMSE_{\Omega_{roi1}}^2 + RMSE_{\Omega_{roi2}}^2 + RMSE_{\Omega_{roi3}}^2},
\end{equation}
where $\Omega_{roi1}$ corresponds to pixels in the vascular region, $\Omega_{roi2}$
correspond to pixels in the tumour region and $\Omega_{roi3}$ corresponds to pixels in
rest of the image domain. %, i.e. $\Omega_{roi3} = \Omega - (\Omega_{roi1} \cup \Omega_{roi2})$.
The RMSE was calculated this way to weigh the small tumour and vascular regions appropriately. In estimating the pharmacokinetic parameters of tissues, obtaining an accurate arterial input function (AIF) is required \cite{Tofts1999}. The AIF can be obtained via population averaging, however, usage of patient specific AIF produces more accurate estimates of the kinetic parameters \cite{Port2001}. The AIF is preferably extracted from an artery feeding the tissues of interest, but it can also be estimated from a venous sinus or vein when an artery is not visible \cite{Lavini2010}. Here the AIF is estimated from the superior sagittal sinus.

\section{Materials and methods}

\begin{figure}[hbtp]
    \centering
    \includegraphics[width=.7\columnwidth]{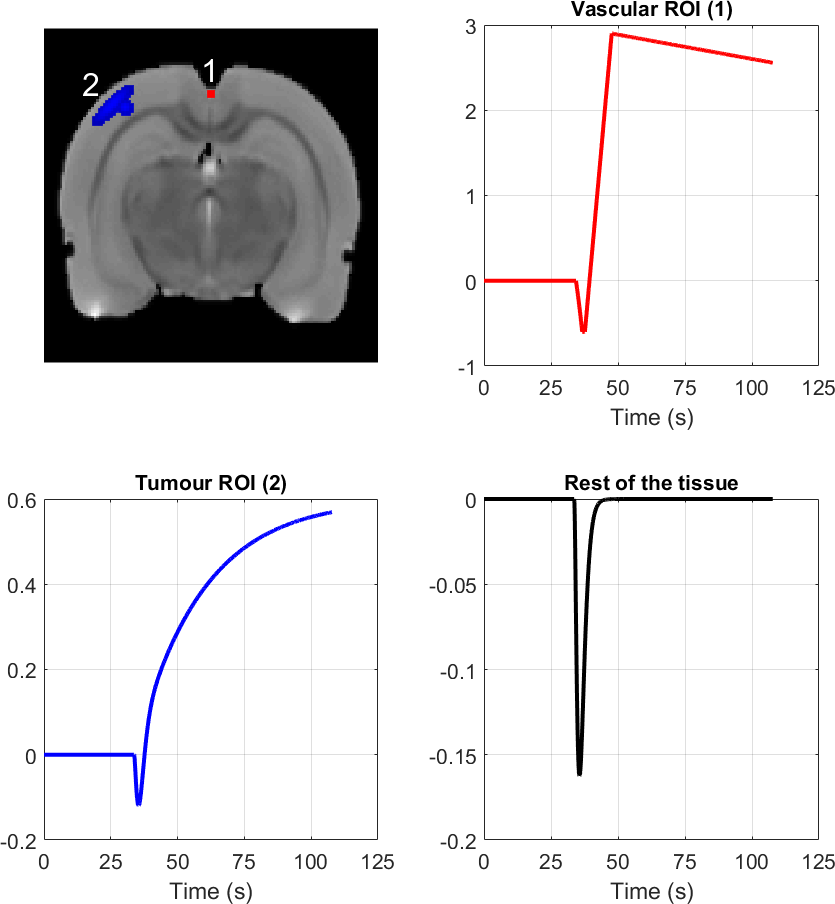}
    \caption{Template signals of different ROIs. Top left: The simulated image with vascular ROI labelled '1' and marked with red and tumour ROI labelled '2' and marked with blue. Top right: The template signal of the vascular ROI. Bottom left: The template signal of the tumour ROI. Bottom right: The template signal of tissue outside the two ROIs. The vertical axis in the three template signals is the multiplier for the signal added to the base signal.}
    \label{fig:simu_templates}
\end{figure}

%\subsection{Simulated data}
\label{sec:simuData}
A simulated test case modelling DCE-MRI measurements of a glioma in rat brain was created. The rat brain phantom was based on the rat brain atlas in \cite{Valdes2011}, and scaled to a size of 128x128. The rat brain image was divided into three subdomains of different signal behaviour: vascular region (labelled '1' and highlighted with red in figure \ref{fig:simu_templates}), tumour region (labelled '2' and highlighted with blue in figure \ref{fig:simu_templates}) and the rest of the brain tissue. The vascular signal region corresponds to the location of the superior sagittal sinus.

A time series of 2800 ground truth images was simulated by multiplying the signal of each pixel with the template of the corresponding region and adding that to the baseline value of the pixels. The tumour signal templates were based on an experimental DCE-MRI measurement described in \cite{Hanhela2019}, where the three different ROIs were identified. Figure \ref{fig:simu_templates} shows the signal templates for each of the different tissue regions.

One spoke of GA k-space data was simulated for each of the simulated images, leading to a dynamic experiment with 2800 spokes of k-space data. The time scale of the simulation was set to be similar to the in vivo measurements in \cite{Hanhela2019} where the 
%repetition time of the GA measurements 
measurement time between consecutive GA spokes was 38.5ms. Gaussian complex noise at 5\% of the mean of the absolute values of the signal was added to the simulated k-space signal. The simulated test case was carried out using a k-space trajectory which combines the golden angle and the concentric squares sampling strategies.

%\subsection{Optimization approach}

In \cite{Hanhela2019} it was found that reconstruction of the form (\ref{eq:spatio_temporal_regularizationL2TV_tReg}) performed optimally with segment length\footnote{Segment length equals the number of radial spokes per image. The number of elements $M$ in the data vector $m_t$ is segment length times number of samples per spoke.} of 34 for a similar data set, thus we selected 34 as the segment length for our reconstructions leading to a temporal resolution of $\sim 1.3$s.

All the regularized reconstructions in this work were computed using the Chambolle-Pock primal-dual algorithm \cite{Chambolle2011}. In the NUFFT implementation of the forward model, the measurements were interpolated into a twice oversampled cartesian grid with min-max Kaiser-Bessel interpolation with a neighbourhood size of 4 \cite{FesslerSutton}.
The regridding reconstructions were computed using a Matlab code developped at Imagerie par r\'{e}sonance magn\'{e}tique m\'{e}dicale et multi-modalit\'{e}s (IR4M) UMR8081, Universit\'{e} Paris-Sud, France.  %\VilleComment{Where is this unit? Refs?}

We remark that when computing the RMS error (\ref{eq:RMSEjoint}), the reconstructed time signals of each pixel were linearly interpolated in the temporal direction to match the temporal resolution of the ground truth phantom. 
%with time resolution of 38.5ms.
%segment length of one (i.e. only spoke of data was simulated for each image in the ground truth sequence).

\section{Results}

\begin{figure}[ht]
  \centering
 \begin{picture}(320,200)
  \put(-15,10){\includegraphics[height=5.5cm]{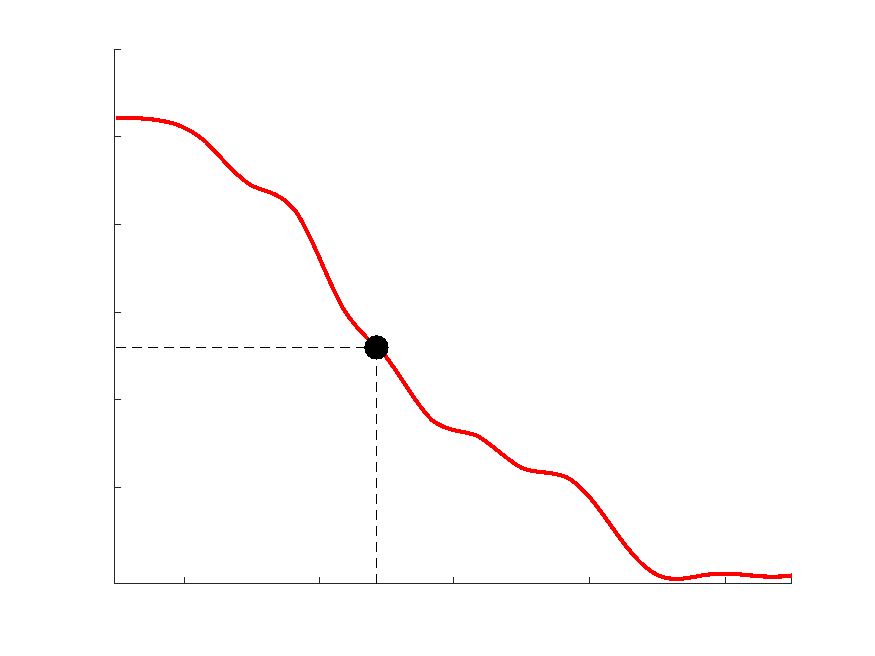}}
 \put(185,23){\includegraphics[height=4.5cm]{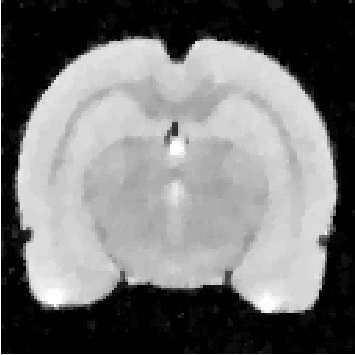}}
 
\put(5, 10){$10^{-7}$}
\put(165, 10){$10^{3}$}
\put(-5,45){ $0.01$}
\put(-1,78){$\hat{S}$}
\put(-5,150){$0.06$}
  \end{picture}
  
  \caption{S-curve for selecting spatial regularization parameter for
    the simulated data case. Left:  Plot of the interpolation curve used to determine the value of
   $\alpha$ such that $S(\alpha) = \hat{S}=0.0259$. 
   Right: Reconstruction (resolution $128\times 128$) of the first
   time frame ($t=1$) using the selected value of $\alpha = 7.3e-4$.}
  \label{fig:Scurve_perf10data}
\end{figure}
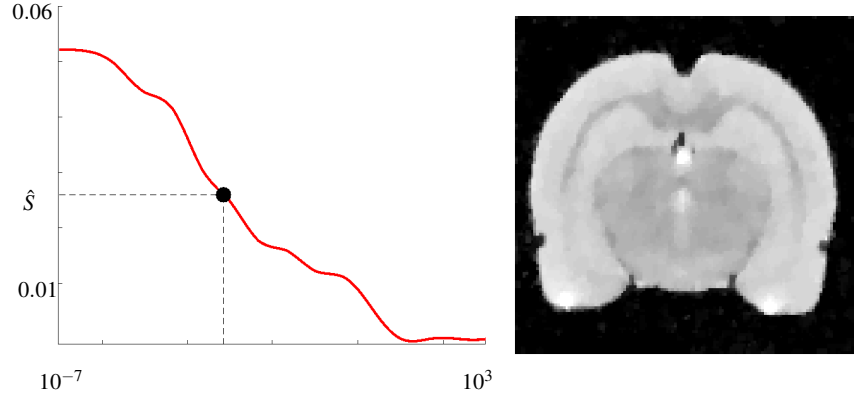

The selection of $\alpha$ was carried out using the first 34 spokes (i.e. the first frame $m_1$) of the DCE-MRI data 
% (baseline measurement)
and then the selected $\alpha$
was used for all the spatio-temporal reconstructions with different temporal regularizations.
The rat brain phantom of section \ref{sec:simuData}  was used to compute the a priori level of
sparsity, i.e. in our case we computed the TV norm of the first time frame of the
dynamic phantom. This resulted in a sparsity level of $\hat{S} = 0.0259$. The
spatial regularization term $\alpha$ was selected using the S-curve as
described in section \ref{sec:Scurve} and resulted in spatial
regularization parameter value of $\alpha = 7.3e-4$. 
The TV norms of the reconstructions for the S-curve were computed with 15 values of alpha ranging on interval $[10^{-7}, 10^3]$. These 15 values of TV norm were then interpolated using Matlab's$^{\textregistered}$ {\it interp} function to 405 values. The resulting S-curve for the determination of $\alpha$ is presented in figure \ref{fig:Scurve_perf10data}. 

In many practical applications, the a priori information, which we use to estimate the value of $\hat{S}$, may come from a different modality or from acquisition with different pulse sequence than the one used in the dynamic measurement. Therefore, in order to compute meaningful estimate of $\hat{S}$ for the TV-regularized case, the reference image has to be scaled such that it is compatible with the measured dynamic data. This normalization of the reference image can be obtained by
\begin{displaymath}
    f_{\mathrm{ref}} = \frac{\Vert m_t \Vert}{\Vert A_t f_{\mathrm{ref}} \Vert}f_{\mathrm{ref}},
\end{displaymath}
where $f_{\mathrm{ref}}$ denotes the reference image, $m_t$
the frame of dynamic data that is used in the S-curve and $A_t$ the respective forward model.

The temporal regularization parameter $\beta$ was selected by computing reconstructions (\ref{eq:spatio_temporal_regularizationL2TV_tReg}) with different values of $\beta$ and then selecting the values which had the smallest joint RMSE error, see (\ref{eq:RMSEjoint}). 

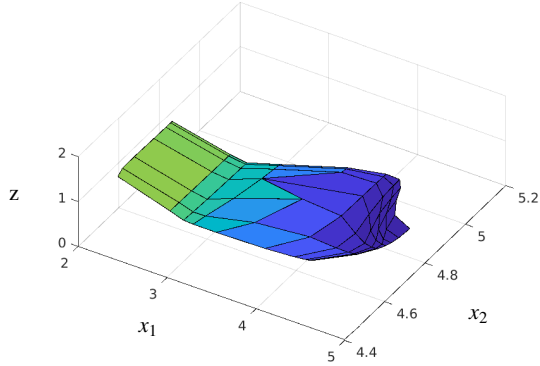
\begin{figure}[ht]
  \centering
 \begin{picture}(320,200)
  \put(45,10){\includegraphics[height=5.5cm]{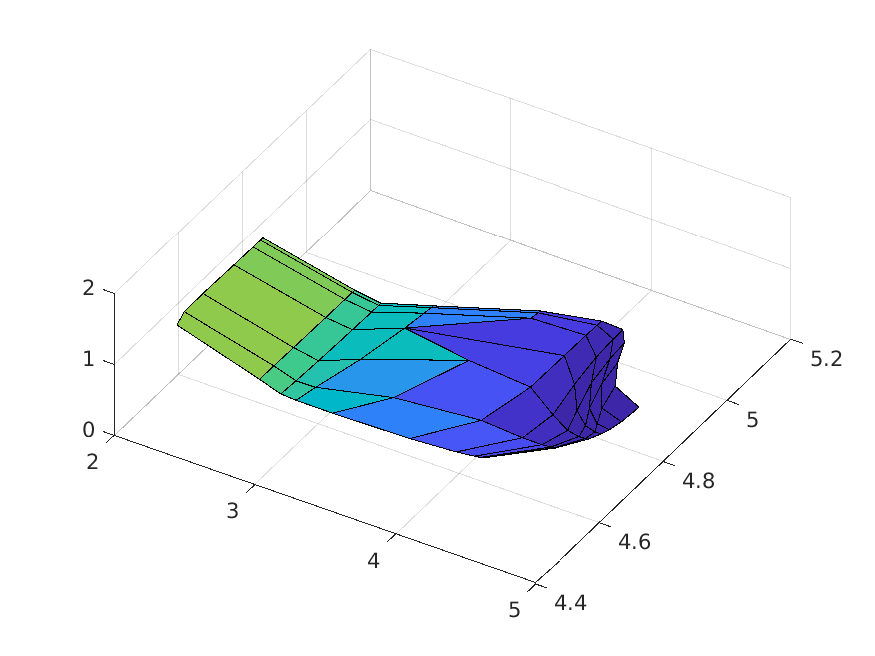}}
  \put(46,80){z}
  \put(95,30){$x_1$}
  \put(220,35){$x_2$}
  \end{picture}
  
  \caption{L-surface for selecting the spatial regularization parameter and the temporal regularization parameter simultaneously, resulting in $\alpha = 3.1623e-4$ and $\beta = 3.1623e-6$. The axes in the image are: $x_1 = \log (\Vert f_{t+1} - f_{t} \Vert)$, $x_2 = \log( \Vert \nabla f_t \Vert_{2,1})$ and $z =  \log (\Vert A_t f_t - m_t \Vert_2)$}.
  \label{fig:Lsurface_perf10data}
\end{figure}

%This process was repeated for all temporal regularization terms. 
This selection resulted in smallest joint RMSE of 0.085 for TS model which corresponded to $\beta = 0.01$, smallest joint RMSE of 0.058 for the TV model corresponding to $\beta = 0.0017$  and smallest joint RMSE of 0.063, corresponding to $\beta = 0.0022$ for the TGV model. As a reference method we selected the regularization parameters $\alpha$ and $\beta$ using L-surface method for the case where we use spatial total variation and temporal total variation as our regularization model.
The resulted L-surface is presented in figure \ref{fig:Lsurface_perf10data}. Application of L-surface on our data resulted parameters $\alpha = 3.1623e-4$ and $\beta = 3.1623e-6$.

\begin{figure}[ht]
  \centering
  \begin{picture}(340,480)
\put(70,400){\includegraphics[width=7.8cm]{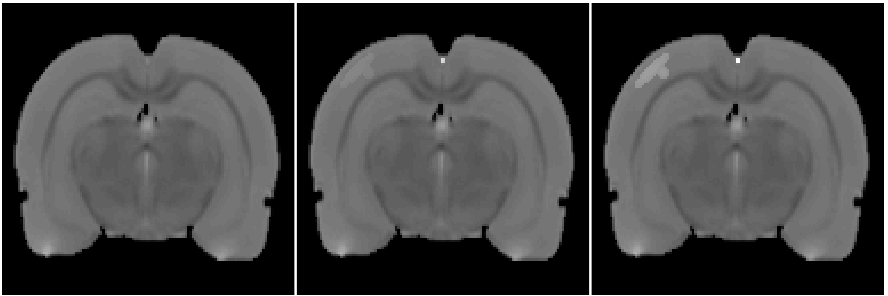}}
\put(70,320){\includegraphics[width=7.8cm]{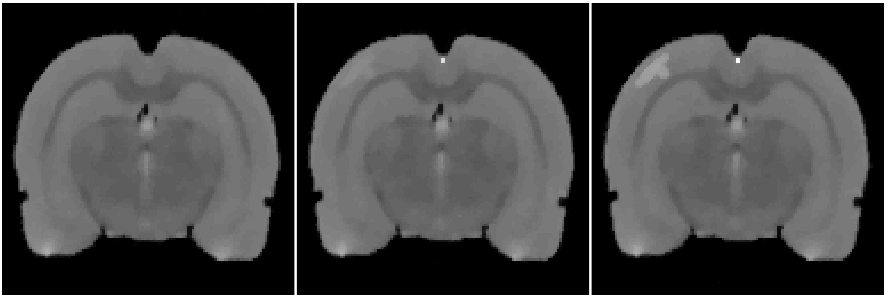}}
\put(70,240){\includegraphics[width=7.8cm]{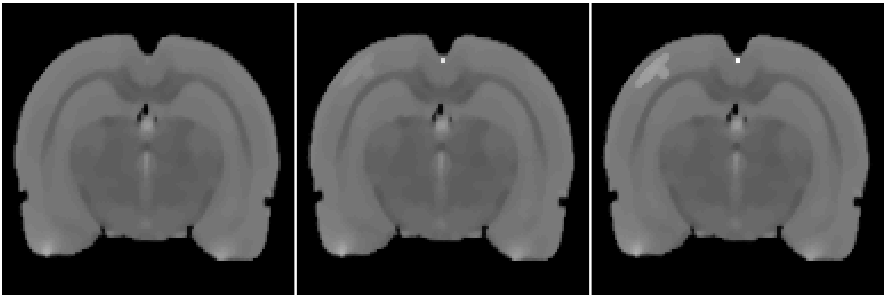}}
\put(70,160){\includegraphics[width=7.8cm]{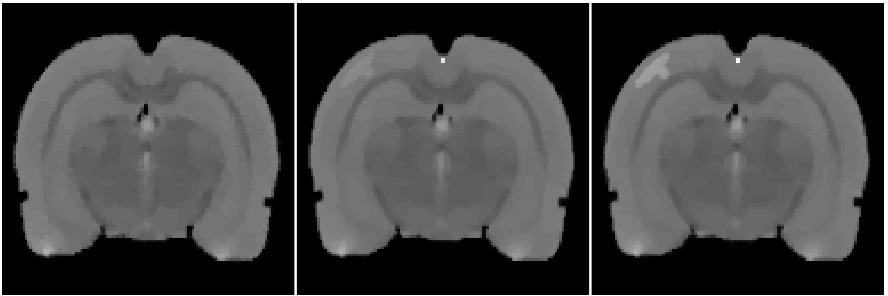}}
\put(70,80){\includegraphics[width=7.8cm]{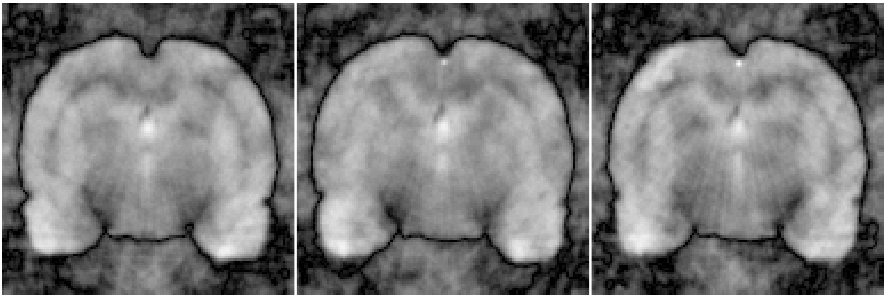}}
\put(70,0){\includegraphics[width=7.8cm]{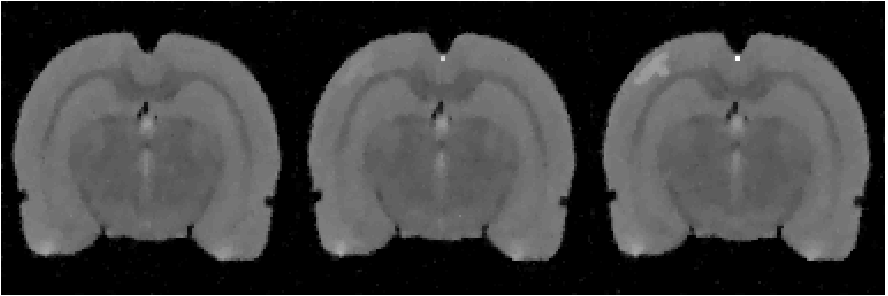}}
%\put(75,158){\textbf{before CE}}
%\put(155,158){\textbf{during CE}}
%\put(235,158){\textbf{after CE}}
\put(30,440){$f_{true}$}
\put(30,360){TS}
\put(30,280){TV}
\put(30,200){TGV}
\put(30,120){Regrid}
\put(30,40){L-surface}
\end{picture}

  \caption{Reconstructions with different temporal regularizations at
    different time frames. From top to bottom: true phantom, TS, TV,
    TGV, regrid and L-surface. From left to right time frames: before, during and after CA administration.}
  \label{fig:Recons3framesSimu}
\end{figure}

\begin{figure}[ht]
  \centering

\setlength{\unitlength}{1mm}
\begin{picture}(117,74)
\put(2,38){\includegraphics[width=36mm]{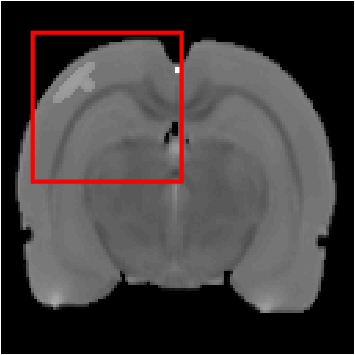}}
\put(40,38){\includegraphics[width=36mm]{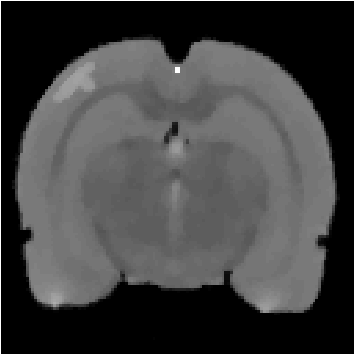}}
\put(78,38){\includegraphics[width=36mm]{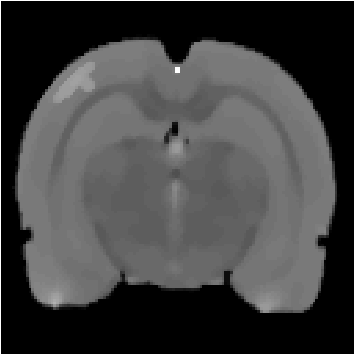}}
\put(2,0){\includegraphics[width=36mm]{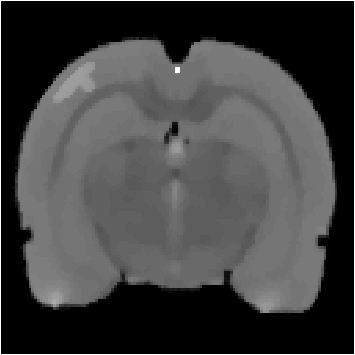}}
\put(40,0){\includegraphics[width=36mm]{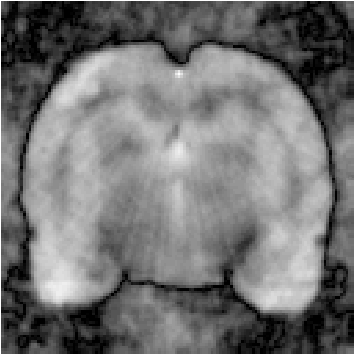}}
\put(78,0){\includegraphics[width=36mm]{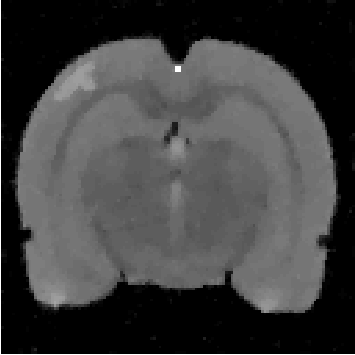}}
\end{picture}

\caption{Reconstructions with different temporal regularizations after CA administration. Top row: true phantom (left), TS reconstruction (middle) and TV reconstruction (right). Bottom row: TGV reconstruction (left),  regrid-reconstruction (middle) and L-surface reconstruction (right). The area highlighted in red is presented as closeups in figure \ref{fig:reconstructionsZOOMED_frame78}.}
  \label{fig:reconstructions_frame78}
\end{figure}

\begin{figure}[ht]
  \centering

\setlength{\unitlength}{1mm}
\begin{picture}(117,74)
\put(2,38){\includegraphics[width=36mm]{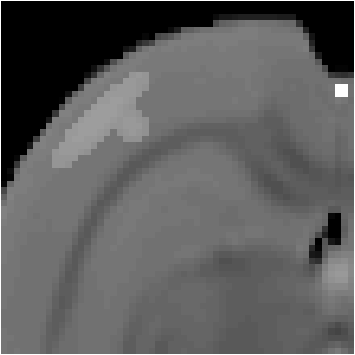}}
\put(40,38){\includegraphics[width=36mm]{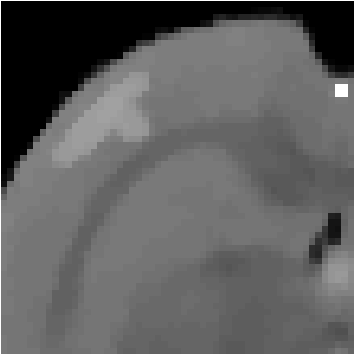}}
\put(78,38){\includegraphics[width=36mm]{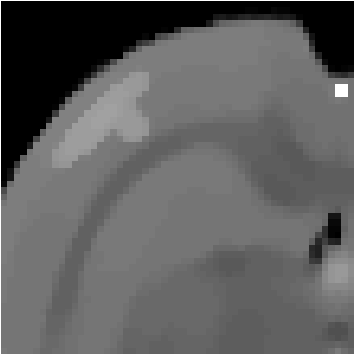}}
\put(2,0){\includegraphics[width=36mm]{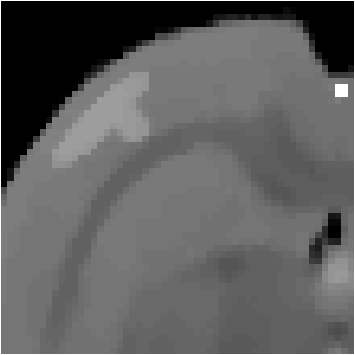}}
\put(40,0){\includegraphics[width=36mm]{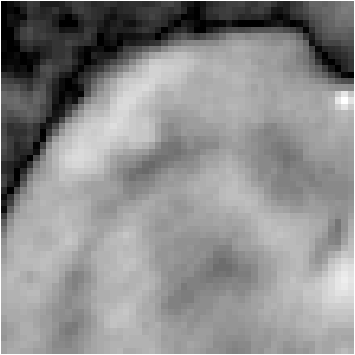}}
\put(78,0){\includegraphics[width=37mm]{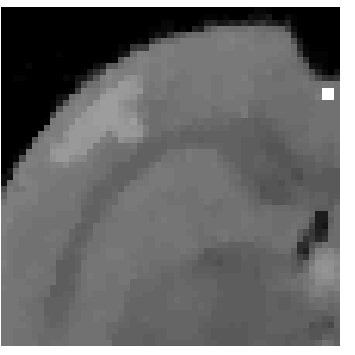}}
\end{picture}

\caption{Closeups of reconstructions with different temporal regularization methods after CA administration. Top row: true phantom (left), TS reconstruction (middle) and TV reconstruction (right). Bottom row: TGV reconstruction (left), regrid-reconstruction (middle) and L-surface reconstruction (right).}
\label{fig:reconstructionsZOOMED_frame78}
\end{figure}

Figure \ref{fig:Recons3framesSimu} shows slices of all reconstructions before, during and after contrast agent administration. 
%Our simulated data represents a brain tumor case with a rodent %specimen. DCE-MRI is useful for the study of such tumors since the %blood-brain-barrier permeability is changed within tumor patients %which causes accumulation of the contrast agent into tumorous tissues %and the tumor becomes visible in the images during and specifically %after the CA administration. 
Figure \ref{fig:reconstructions_frame78} shows the reconstructed images with different methods for one time frame. The top row of figure \ref{fig:reconstructions_frame78} shows the phantom with a red square denoting a domain that is presented as a closeup in figure \ref{fig:reconstructionsZOOMED_frame78}  for all of the reconstructions. 

As can be seen from figures \ref{fig:Recons3framesSimu} - \ref{fig:reconstructionsZOOMED_frame78}, the classical regridding method fails on such high time resolution data as employed here, and thus 
%This can be expected due to the lack of regularization 
we leave out the regridded reconstruction from the figures of the temporal evolution of the ROI signals. However the L-surface method seems to work nicely on our data, so we include it in the temporal evolution studies. The temporal evolutions in the vascular domain and in the tumor region are averages of $\Omega_{roi1}$ and $\Omega_{roi2}$, respectively.

\begin{figure}[ht]
  \centering
  \begin{picture}(340,200)
\put(0,10){\includegraphics[width=5cm]{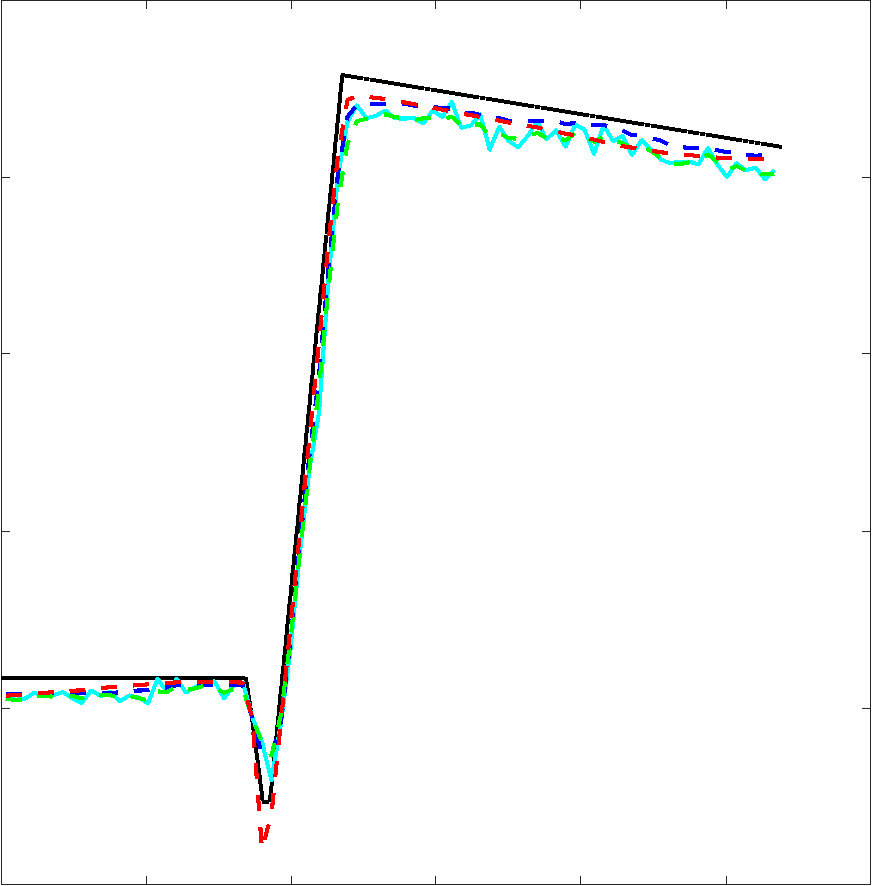}}
\put(170,10){\includegraphics[width=5cm]{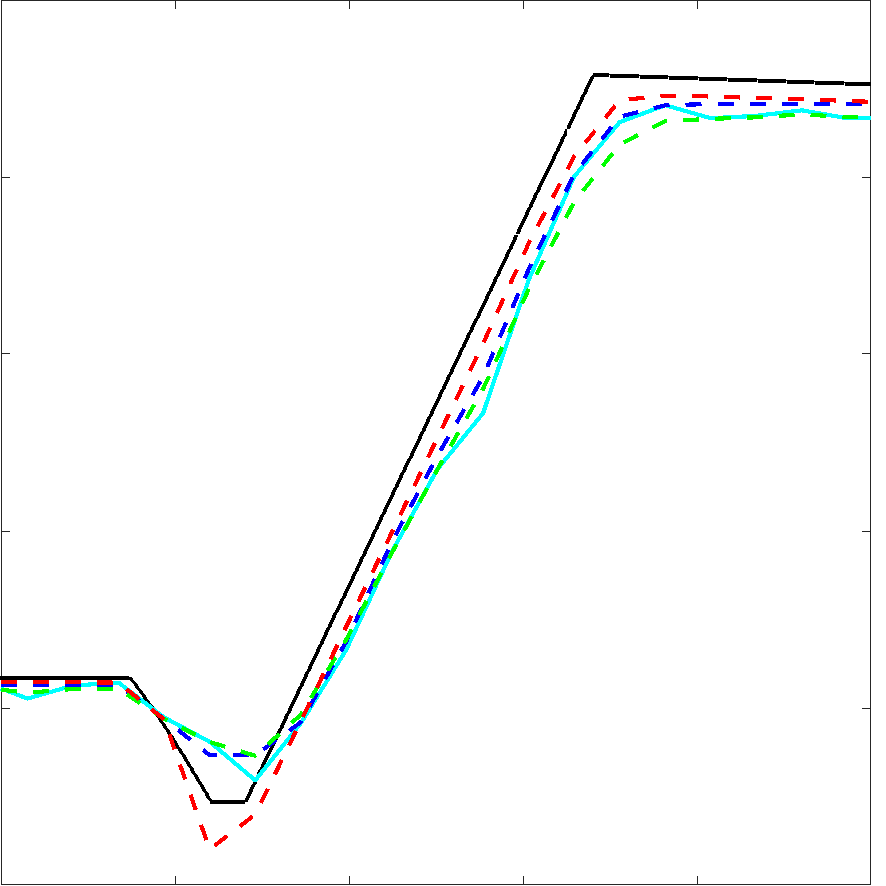}}
\end{picture}

  \caption{Reconstructions of vascular region ($\Omega_{roi1}$) with the different methods at their optimal parameters
according to the joint RMSE. Black line: true signal, blue line: TV
reconstruction, green line: TS reconstruction, red line: TGV
reconstruction and light blue line: L-surface reconstruction. Left: temporal evolution during all time
frames. Right: closeup image during CA administration.}
  \label{fig:Artery_signal_plots}
\end{figure}

\begin{figure}[ht]
  \centering
  \begin{picture}(340,200)
\put(0,10){\includegraphics[width=5cm]{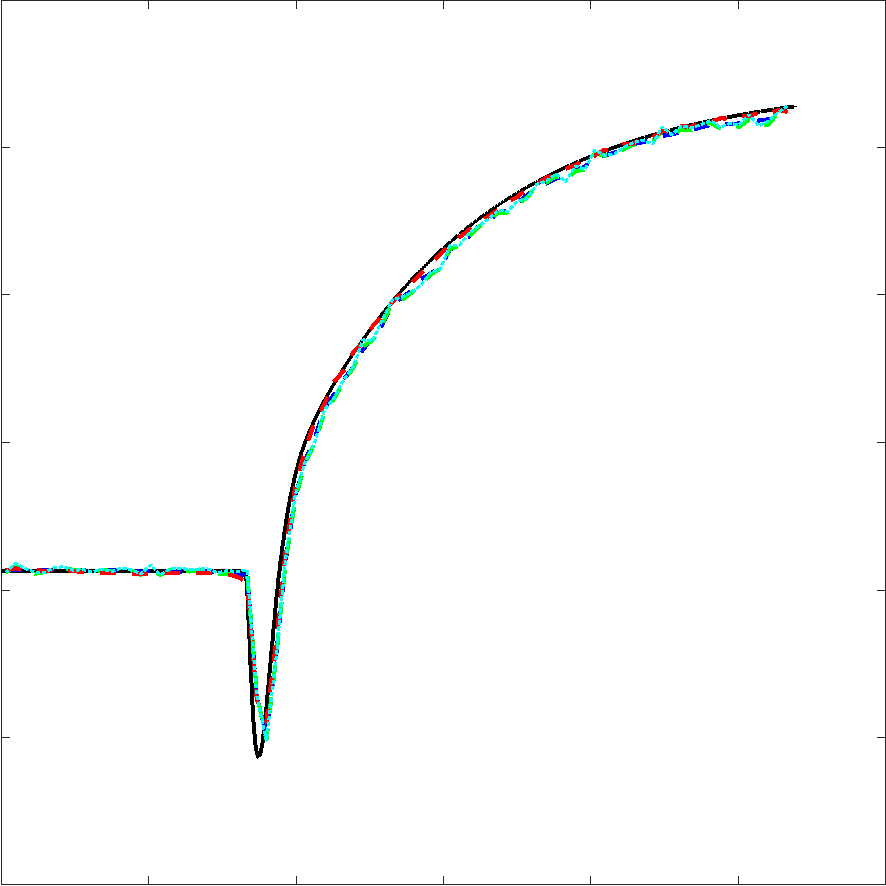}}
\put(170,10){\includegraphics[width=5cm]{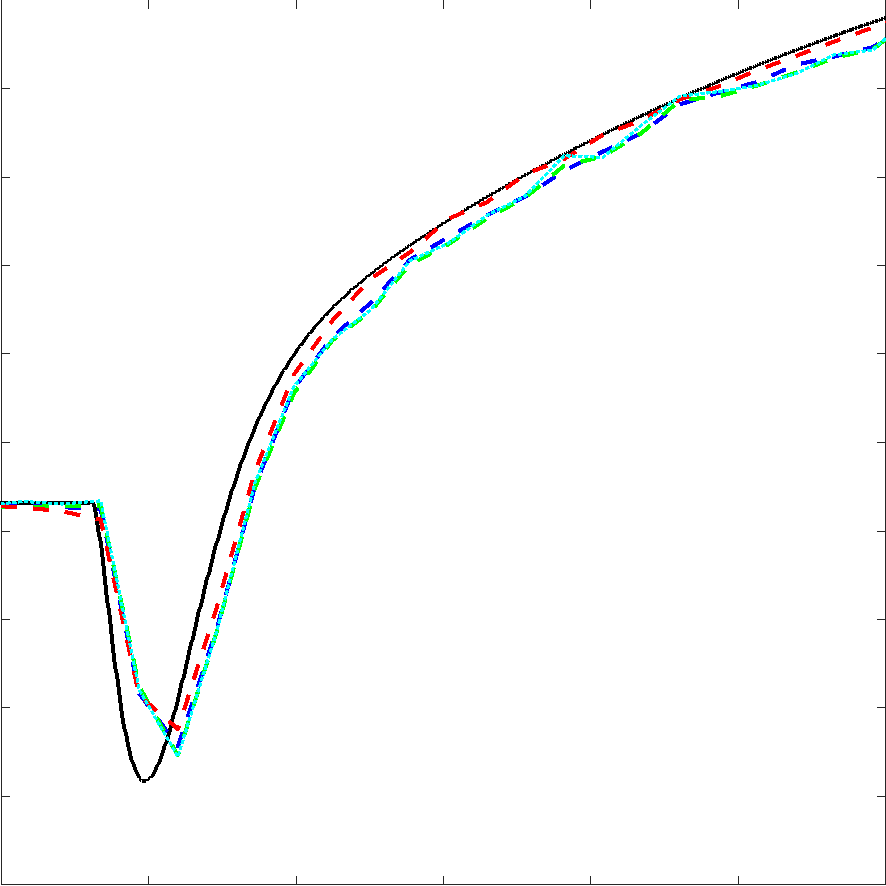}}
\end{picture}

\caption{Reconstructions of tumor region ($\Omega_{roi2}$) with the different methods at their optimal parameters according to the joint RMSE. Black line: true signal, blue line: TV
reconstruction, green line: TS reconstruction, red line: TGV
reconstruction and light blue line: L-surface reconstruction. Left: temporal evolution during all time
frames. Right: closeup image during CA administration.}
  \label{fig:Tumor_signal_plots}
\end{figure}

Figure \ref{fig:Artery_signal_plots} shows the time signals of the reconstructions in the vascular region ($\Omega_{roi1}$) with the different temporal
regularization models.
%using $\alpha$ as selected by the S-curve
%method and optimal value of $\beta$ according to RMSE. 
Corresponding signals of the reconstructions in the tumor region (i.e. in $\Omega_{roi2}$) are shown in figure \ref{fig:Tumor_signal_plots}. The tumor region is accurately reconstructed with all the methods, with only small differences between the methods, L-surface method being the noisiest. In the vascular region, the methods have some differences with TGV having the best reconstruction quality and TS having the worst reconstruction quality. The TS method shows smoothing at both the maximum and minimum signal levels whereas TGV reconstructs the fast signal change of the vascular region most reliably. L-surface method is again more noisy than the other reconstructions in the vascular region and its performance of reconstructing the vascular signal falls between TS and TV most likely due to the temporal regularization parameter being too small whereas the spatial regularization parameter is on the same order of magnitude as the parameter obtained with S-curve.

\section{Conclusions}

Variational regularization based solutions for dynamic MRI problems usually include two regularization parameters, one for the spatial and one for the temporal regularization, that the user has to select. Typically, the selection of both of the parameters is carried out manually based on visual assessment of the reconstructed images. 
In this work we proposed to use the S-curve method for the automatic selection of the spatial regularization parameter, leaving the time regularization parameter the only free parameter. The S-curve method selects the regularization parameter based on the expected sparsity of unknown images in domain of the regularization functional. Furthermore, the method requires computation of the reconstructions with relatively few values of the parameter, making it computationally efficient. 
The approach was demonstrated to lead to a feasible choice of the spatial regularization parameter in a simulated DCE-MRI experiment of rat brain. 
The reconstructions were also computed with three different temporal regularization models with the same fixed spatial regularization parameter, demonstrating the robustness of the approach with different time regularization models.

While we proposed automatic selection of the spatial regularization parameter,
the temporal regularization parameter was still selected manually. In this work, we selected the temporal regularization parameter by computing RMS errors with respect to a ground truth. If on the other hand the ground truth is unknown, the temporal regularization parameter could be selected based on visual assessment of reconstructed images, leaving $\beta$ to be the only free parameter. In the future work, we aim to study methods for automatic selection of the time regularization parameter as well. One possibility could be to extend the S-curve to select a parameter which leads to an expected sparsity level in the domain of the temporal regularization. A feasible estimate for the expected level of
sparsity in the time direction could potentially be extracted from the changes in the dynamic measurement data.

\section*{Acknowledgments}
This work  was supported by Jane and Aatos Erkko foundation and the Academy of Finland, Centre of Excellence in Inverse Modelling and imaging (project 312343). \\
%{\sc All: Please add relevant numbers and projects}
%\section*{References}
\bibliographystyle{plain}
\bibliography{Inverse_problems_references.bib}

\end{document}